\documentclass{moriond}

\usepackage{amsmath}
\usepackage{amssymb}
\usepackage[dvipsnames]{xcolor}

\def\Journal#1#2#3#4{{#1} {\bf #2}, #3 (#4)}

\def\PRD{{\em Phys. Rev.} D}

\def\aap{\em Astronomy \& Astrophysics}
\def\app{\em Astroparticle Physics}

\def\cqg{\em Classical \& Quantum Gravity}
\def\IJMPD{\em Int. J. Mod. Phys. D}
\def\LRR{\em Living Reviews in Relativity}
\def\mnras{\em Mon. Notices Royal Astron. Soc.}

\def\be{\begin{equation}}
\def\ee{\end{equation}}
\def\bea{\begin{eqnarray}}
\def\eea{\end{eqnarray}}

\def\LCDM{$\Lambda$CDM}

\definecolor{darkgreen}{cmyk}{0.85,0.2,1.00,0.05}

\newcommand{\Photo}{}

\begin{document}
\vspace*{4cm}
\title{ON THE ROAD TO PERCENT ACCURACY: THE REACTION WAY}

\author{MATTEO CATANEO}

\address{Argelander-Institut für Astronomie, Auf dem Hügel 71, D-53121 Bonn, Germany}

\maketitle

\abstracts{
% This is where the abstract should be placed. It should consist of one paragraph
% and give a concise summary of the material in the article below.
% Replace the title, authors, and addresses within the curly brackets
% with your own title, authors, and addresses; please use
% capital letters for the title and the authors. You may have as many authors and
% addresses as you wish. It's preferable not to use footnotes in the abstract
% or the title; the
% acknowledgments for funding bodies etc. are placed in a separate section at
% the end of the text.
Taking advantage of the unprecedented statistical power of upcoming cosmic shear surveys will require exquisite knowledge of the matter power spectrum over a wide range of scales. Analytical methods can achieve such precision only up to quasi-linear scales. For smaller non-linear scales we must resort to $N$-body and hydrodynamical simulations, which despite recent technological advances and improved algorithms remain computationally expensive. Over the past decade machine learning and the advent of emulators have propelled our ability to hit the target accuracy with impressive computing efficiency. Yet, realistically these techniques will be able to produce high-precision non-linear matter power spectra only for a restricted sub-set of extensions to the ``vanilla'' \LCDM{} cosmology.
I will present a promising alternative to alleviate these shortcomings that draws strength from the combination of halo model, perturbation theory and emulators--the \emph{reaction} framework. I will show how a power spectrum evolved in the standard cosmology can be readily adjusted to account for physics beyond \LCDM{}, and discuss the accuracy of the reaction for well-known modifications to gravity, dark energy parametrizations and massive neutrino cosmologies.}

\section{Introduction}

The last two decades have seen the rise of the $\Lambda$-Cold Dark Matter (\LCDM{}) paradigm to Standard Model of Cosmology--a status granted for its ability to explain with only a handful of parameters a diverse array of observations tracing a staggering 13.8 billion years of cosmic evolution (for recent analyses see, e.g., Planck collaboration~\cite{Planck18}, eBOSS collaboration~\cite{eBOSS}, DES~\cite{DESY3} and KiDS~\cite{KiDS1000} collaborations). The ever-increasing volume of data continues to challenge the \LCDM{} model, and tantalizing tensions are now emerging between early- and late-time cosmological probes~\cite{DiValentino21a,DiValentino21b} potentially exposing cracks in the theory. When taken together with our largely phenomenological (rather than fundamental) understanding of the content of the Universe, this fact encourages the exploration of alternative models in an attempt to find (most likely, constrain) deviations from the standard picture. 

Central to many cosmological analyses is the 2-point correlation function of the matter density field, or equivalently its Fourier transform, the matter power spectrum. In the era of Stage IV surveys, accurate predictions for this statistic down to small non-linear scales will be essential to derive tight as well as robust constraints on parameters informing us whether new physics is at work~\cite{Heymans&Zhao18,Taylor18}, be it yet unknown interactions, undetected particles or particle properties. As the clustering of matter can react quite dramatically to changes to General Relativity (GR), testing gravity on cosmological scales is high up on the agenda of all science collaborations~\cite{DESY1:extended,Troester21,Amendola18,Ishak19}. Modelling the non-linear power spectrum in modified gravity scenarios is the focus of this talk, but I will also discuss how our framework can capture the impact of massive neutrinos--particularly relevant for its degeneracy with fifth force effects~\cite{Baldi14}--and evolving dark energy (i.e., $w \neq -1$).

\section{Non-linear matter power spectrum predictions for \LCDM{} extensions}

$N$-body simulations are the gold standard for the study of non-linear structure formation, and they provide the means to directly estimate the statistical properties of the matter density field, including the power spectrum. On the flip side is their considerable computational cost. 
%with a single run requiring no less than several hundred node-hours on high-performance computing infrastructures even for nowadays fastest codes~\cite{Knabenhans21,Maksimova21}. 
Unfortunately, the complexities of new physics 
%(e.g., non-minimally coupled scalar fields) 
in \LCDM{} extensions only worsen this problem~\cite{Llinares18}. Methods to predict the non-linear power spectrum based on (semi-)analytical or machine learning approaches can provide a fast alternative to simulations, but their validity is restricted to particular domains (e.g., quasi-linear regime, certain class of cosmologies etc.)
%: perturbation theory works only on large quasi-linear scales~\cite{}; the inaccuracies of the halo model on intermediate scales are difficult to remove~\cite{}; the evaluation of fitting functions is straightforward but their applicability and generality to future data sets is questionable~\cite{}; although emulators have proven to be both accurate and computationally efficient~\cite{}, predictions are inherently limited to the class of cosmologies used in the training phase.  

By capitalizing on the strengths of well-established techniques (perturbation theory, halo model and emulators), the \emph{reaction} method described below provides a framework for predicting the non-linear matter power spectrum of a broad class of cosmologies while meeting the accuracy requirements set by Stage IV surveys. Moreover, its small computation time is an especially attractive feature for likelihood analyses.

\section{The Reaction framework}

In this framework the non-linear matter power spectrum of our target cosmology (which we call the `real' cosmology) can be obtained from that of a particular reference cosmology (the `pseudo' cosmology) as~\cite{Cataneo19}
\begin{equation}
    P^{\rm real}(k,z) = \mathcal{R}(k,z) \times P^{\rm pseudo}(k,z) \, ,
\end{equation}
where $\mathcal{R}$ is a scale- and time-dependent quantity called \emph{reaction}. The linear growth of the pseudo cosmology follows that of a flat and massless neutrino \LCDM{} cosmology, with the important difference that the shape and amplitude of its linear power spectrum are specified by the real cosmology, i.e.
\begin{equation}\label{eq:pseudo_def}
    P_{\rm L}^{\rm pseudo}(k,z) = P_{\rm L}^{\rm real}(k,z) \, .
\end{equation}
%
%For any final redshift, $z$, this equality can always be satisfied if at some initial time deep in the matter-dominated era, $z_{\rm i}$, we have
%%
%\begin{equation}
%    P_{\mathrm{L}}^{\rm {pseudo}}\left(k, z_{\mathrm{i}}\right)=\left[\frac{D_{\Lambda}\left(z_{\mathrm{i}}\right)}{D_{\Lambda}\left(z\right)}\right]^{2} %P_{\mathrm{L}}^{\mathrm{real}}\left(k, z\right) \, ,
%\end{equation}
%%
%with $D_\Lambda$ denoting the \LCDM{} linear growth. Recall that in a flat and massless neutrino \LCDM{} cosmology the only parameter controlling the linear growth is the total matter density, $\Om$, and for simplicity here we use $\Om^{\rm pseudo} = \Om^{\rm real}$. Note that in general the growth of structure in non-standard cosmologies (the `real' cosmologies) differs from that of \LCDM{}, therefore for the initial conditions we have $P_{\rm L}^{\rm pseudo}(k,z_{\rm i}) \neq P_{\rm L}^{\rm real}(k,z_{\rm i})$. 
The reason for choosing the pseudo cosmology in such a way is twofold: (i) it ensures $\mathcal{R} \rightarrow 1$ on linear scales; (ii) since to a good approximation the spherical collapse threshold is only weakly dependent on cosmology, from the Press-Schechter formalism we expect the halo mass functions of the pseudo and real cosmology to be rather similar. We shall see that it is this similarity that enables us to predict the reaction with the required accuracy. 

To take advantages of the complementarity between different techniques, we use flexible semi-analytical methods for the reaction together with an accurate simulation-based route for the pseudo cosmology non-linear power spectrum, so that errors in the predicted beyond-\LCDM{} non-linear power spectrum will be mostly associated with inaccuracies in the reaction modelling. The challenge consists in computing the reaction at percent level for all wavenumbers $k \lesssim 10 \, h/{\rm Mpc}$, and at the same time efficiently evaluate $P^{\rm pseudo}(k,z)$ for a broad class of linear power spectrum shapes.

\subsection{Pseudo cosmology emulator}

%\begin{figure}
%\centerline{\includegraphics[width=0.5\linewidth]{Plot_GPAcc_PhysMods.png}}
%\caption[]{Prediction accuracy of our pseudo-cosmology emulator (magenta band) for a range of viable models including $f(R)$ gravity, massive neutrinos %and their combinations. Figure taken from Giblin et al. (2020)~\cite{}.}
%\label{fig:emulator}
%\end{figure}

%In recent years emulators of the non-linear matter power spectrum have become ever more accurate and efficient, and various computational efforts have pushed their boundaries in the vast theory landscape~\cite{}. By nature machine learning algorithms require a training set, which in this particular case consists of a collection of non-linear power spectra for carefully sampled cosmologies. These power spectra are measured from $N$-body simulations, and just for the most common \LCDM{} extensions we need $\sim$250 simulations to achieve better than percent accuracy in the emulator predictions~\cite{}. Once we start considering more exotic cosmologies this approach quickly becomes prohibitive. In fact, the sharp increase in the number of required simulations and the additional computing load per simulation (i.e., we must solve for new physics) implies we can build emulators for only a handful of non-standard cosmologies. 

Thanks to the concept of pseudo cosmology the reaction framework can substantially reduce our reliance on beyond-\LCDM{} simulations, therefore placing the construction of a wide scope emulator within the realm of possibility. The advantage is that the computing time for a pseudo cosmology simulation is similar to that for a \LCDM{} run. However, compared to conventional emulators the pseudo cosmology emulator must capture a variety of linear power spectrum shapes that cannot be reproduced by the standard cosmological parameters. Giblin et al. (2019)~\cite{Giblin19} developed a proof-of-concept emulator precisely with this idea in mind. In addition to the five \LCDM{} parameters they introduced a new set of effective parameters, $\Delta\alpha_{1{\rm -}8}$, describing the ratio $S(k) \equiv P_{\rm L}^{\rm pseudo}(k)/P_{\rm L}^{\Lambda}(k)$ for a large class of cosmologies converging to \LCDM{} on linear scales. 
%Each $\Delta\alpha_i$ weighs a function, $\Phi_i(k)$, such that 
%%
%\begin{equation}
%    S(k) \approx 1+\sum_{i=1}^{n_{\Phi}} \Phi_{i}(k) \Delta \alpha_{i} \, ,
%\end{equation}
%%
%where $n_{\Phi}$ is the number of functions included in the shape reconstruction. 
%Principal Component Analysis on a set of random shapes provided the functions $\Phi_i(k)$, and they found that $n_\Phi = 8$ is enough to reach a reconstruction accuracy $\lesssim 0.5 \%$ for $f(R)$ gravity + massive neutrino cosmologies. 
By using Halofit as a proxy for the $N$-body simulations, they constructed a 13-dimensional Gaussian Process emulator trained on 500 parameter combinations distributed in a Latin Hypercube, and showed that this initial setup can already predict the pseudo non-linear power spectrum of their test cosmologies at 2\% level. 

\subsection{The reaction}

\begin{figure}
\begin{minipage}{0.5\linewidth}
\centerline{\includegraphics[width=0.9\linewidth]{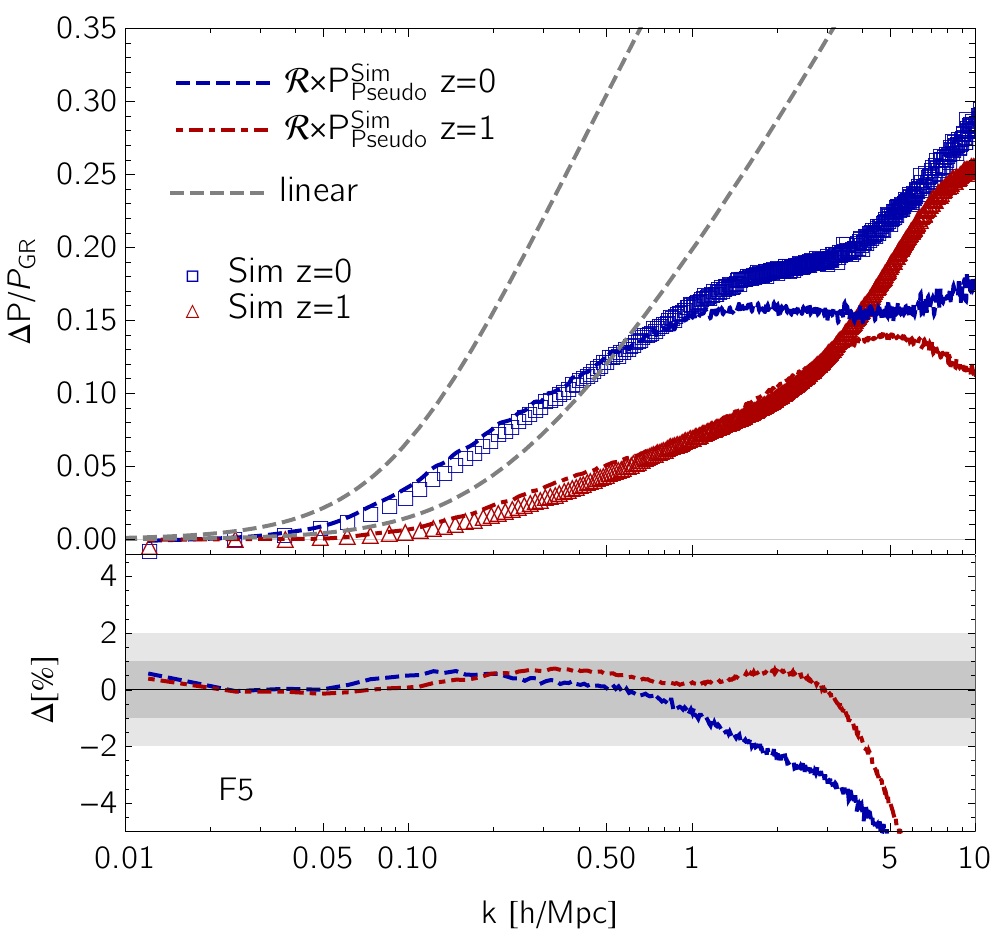}}
\end{minipage}
\hfill
\begin{minipage}{0.5\linewidth}
\centerline{\includegraphics[width=0.9\linewidth]{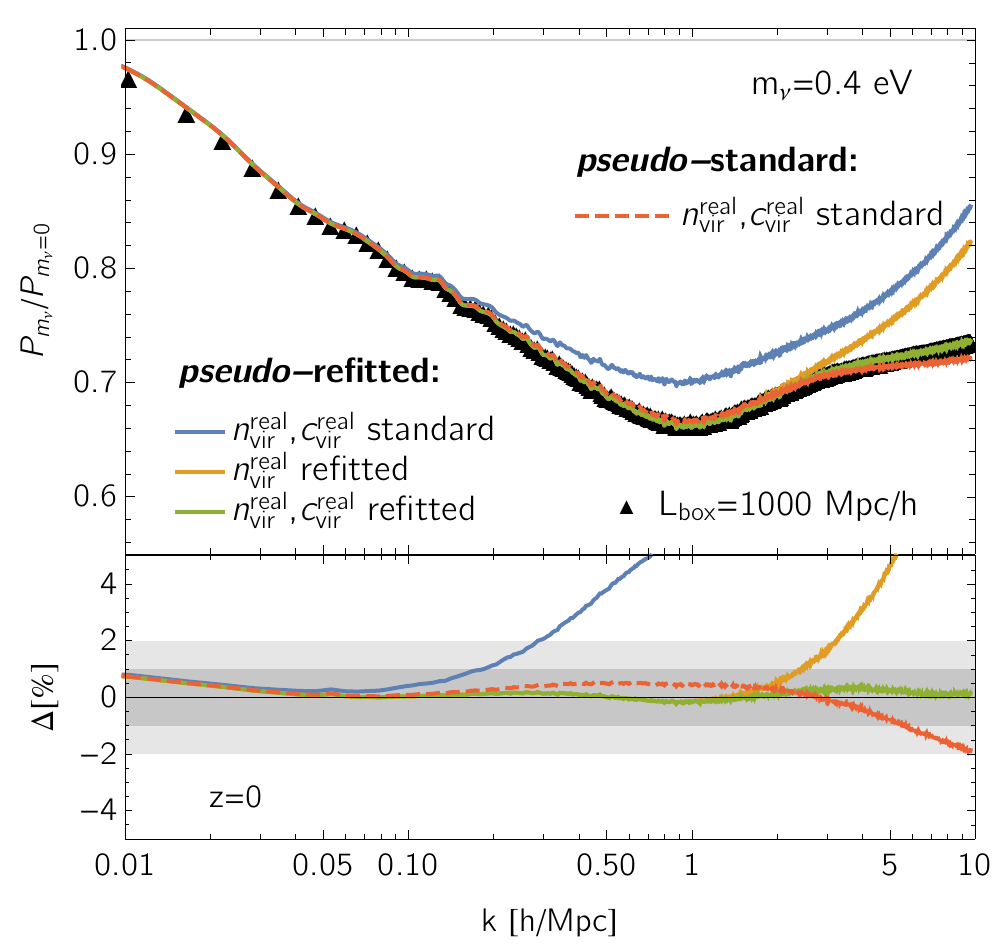}}
\end{minipage}
\caption[]{\emph{Left:} Matter power spectrum fractional enhancement in $f(R)$ gravity compared to GR. Symbols corresponds to the simulations and lines are our predictions based on the reaction. The dashed grey lines follow from linear theory calculations. The bottom panel shows that our predictions are within 1\% of the simulations for $k \lesssim 1 \, h/{\rm Mpc}$. From Cataneo et al. (2019)~\cite{Cataneo19}. \emph{Right:} matter power spectrum ratio of a massive neutrino to a massless neutrino cosmology. The black triangles are simulation measurements. If the pseudo cosmology halo mass function and halo profiles extracted from the simulations are used to predict the reaction, we see that the inclusion of the corresponding measurements for the real cosmology brings the predictions in excellent agreement with the simulations (from blue to orange to green). From Cataneo et al. (2020)~\cite{Cataneo20}.}
\label{fig:reaction_predictions}
\end{figure}

To model the reaction we resort to the halo model (HM) and standard perturbation theory (SPT). For modified gravity and dark energy cosmologies with a fraction, $f_{\nu}$, of the total matter density (m) in massive neutrinos ($\nu$) we have~\cite{Bose21}
\begin{equation}\label{eq:reaction}
    \mathcal{R}(k,z) \approx \frac{P_{\rm HM}^{\rm real}(k,z)}{P_{\rm HM}^{\rm pseudo}(k,z)} =\frac{\left(1-f_{\nu}\right)^{2} P_{\mathrm{HM}}^{(\mathrm{cb})}(k,z)+2 f_{\nu}\left(1-f_{\nu}\right) P_{\mathrm{HM}}^{(\mathrm{cb\nu})}(k,z)+f_{\nu}^{2} P_{\mathrm{L}}^{(\nu)}(k,z)}{P_{\mathrm{L}}^{(\mathrm{m})}(k,z)+P_{1 \mathrm{h}}^{\mathrm{pseudo}}(k,z)} \, ,
\end{equation}
%
%with
%%    
%\begin{eqnarray}
%    &P_{\mathrm{HM}}^{(\mathrm{cb} \nu)}(k,z)& \approx \sqrt{P_{\mathrm{HM}}^{(\mathrm{cb})}(k,z) P_{\mathrm{L}}^{(\nu)}(k,z)} \, ,  \\
%    &P_{\mathrm{HM}}^{(\mathrm{cb})}(k,z)& = \left[(1-\mathcal{E}) e^{-k / k_{\star}}+\mathcal{E}\right] P_{\mathrm{L}}^{(\mathrm{cb})}(k,z)+P_{1 %\mathrm{h}}^{(\mathrm{cb})}(k,z) \, ,
%\end{eqnarray}
%%
with $P_{\mathrm{HM}}^{(\mathrm{cb} \nu)}(k,z)$ and $P_{\mathrm{HM}}^{(\mathrm{cb})}(k,z)$ being, respectively, the cross power spectrum of the massive neutrinos and the CDM+baryons (cb), and the predicted halo model non-linear power spectrum for the CDM+baryons component. 
%The quantities $\mathcal{E}$ and $k_\star$ are derived from the halo model and SPT (see Cataneo et al. 2019~\cite{Cataneo19} for details), and 
To lighten the notation the designation `real' has been omitted for all terms in the numerator of Eq.~\ref{eq:reaction}. The expressions for the pseudo and real 1-halo terms
%, $P_{1 \mathrm{h}}^{\mathrm{pseudo}}$ and $P_{1 \mathrm{h}}^{(\mathrm{cb})}$, 
follow the standard halo model prescription with halo mass functions and halo profiles adapted to the specific cosmologies. %Eq.~\ref{eq:reaction} reduces to simpler forms when extensions to the standard cosmology are considered one at the time~\cite{Cataneo19,Cataneo20}. 
The left panel of Fig.~\ref{fig:reaction_predictions} shows that when the halo profiles are not fully corrected for fifth force effects the reaction predictions can match well the simulation measurements up to $k \sim 1 \, h/{\rm Mpc}$. The right panel, instead, illustrates the strong connection enabled by the reaction framework between the halo properties (abundance and profiles) and the matter power spectrum: after calibrating the halo mass functions of the pseudo and real cosmologies with simulations, our power spectrum predictions can achieve sub-percent accuracy up to $k \sim 1 \, h/{\rm Mpc}$ (orange line); including also information from the halo profiles improves the agreement with the simulations deep in the non-linear regime (green line). 
%This feature is in stark contrast to the standard halo model predictions~\cite{}, and boils down to the likeness between the real and pseudo cosmology halo mass functions. 

Recently Bose et al. (2021)~\cite{Bose21} have extended the reaction framework to include a class of interacting dark energy cosmologies, and packaged this formalism in \texttt{ReACT}, a publicly available C++ code and Python wrapper valuable for likelihood analyses of cosmic shear data~\cite{Bose20}.

\section{Future directions}

This research programme can be expanded in many ways: (\emph{i}) the reaction framework uses the same halo mass function to predict both the matter power spectrum and the abundance of massive halos. This link has the potential to break parameter degeneracies in joint analyses of cosmic shear data and cluster number counts; (\emph{ii}) our formalism is adaptable to a larger class of dark energy and modified gravity cosmologies falling under the umbrella of the Horndeski's theory. By generalizing the spherical collapse dynamics we will be able to capture the phenomenology of these models too. (\emph{iii}) AGN feedback redistributes the gas content of massive halos ($M_{\rm h} \gtrsim 10^{13.5} \, M_{\odot}/h$) thus changing their overall density profiles. This effect can be readily incorporated into the reaction formalism, which together with multi-wavelength observations of groups and clusters of galaxies will put us in a better position to break degeneracies between astrophysical systematics and cosmology, currently a major limiting factor in cosmic shear analyses; (\emph{iv}) the reaction framework enables the simultaneous treatment of various physics beyond the standard gravity-only paradigm, and thanks to the growing library of models implemented in \texttt{ReACT} it will be easier in the future to study the combined effect of exotic dark matter particles, new interactions, gravity beyond GR and astrophysics on the non-linear matter power spectrum; (\emph{v}) finally, Cataneo et al. (2021)~\cite{Cataneo21} found that for sufficiently large smoothing scales the real and pseudo modified gravity probability distribution functions of the matter density field are remarkably similar. This bodes well for potential applications of the reaction to higher-order statistics, e.g. the matter bispectrum, which will be essential to access the non-Gaussian information stored in the large-scale structure. 

\section*{Acknowledgments}

A big thanks to the organisers of the $56^{\rm th}$ Rencontres de Moriond for giving us some relief from two years of virtual meetings with this fantastic in-person conference. The reaction framework rests on the amazing work of my collaborators, and to them goes my deepest gratitude. The author is supported by a Research Fellowship of the Alexander von Humboldt Foundation.

\end{document}